\documentclass[aps,twocolumn,floatfix,showpacs,amsmath,amsfonts]{revtex4}

\usepackage{graphicx}
\usepackage{bm}
\newcommand{\z}{{\bf z}}
\newcommand{\y}{{\bf y}}
\newcommand{\x}{{\bf x}}
\newcommand{\p}{{\bf p}}

\newcommand{\g}{\gamma}
\newcommand{\uu}{{\bf u}}

\newcommand{\pa}{\partial}
\newcommand{\lela}{\left\langle}
\newcommand{\rira}{\right\rangle}
\renewcommand{\chi}{\psi}

\begin{document}

\title{Fictitious time wave packet dynamics:
II. Hydrogen atom in external fields}

\author{Toma\v{z} Fab\v{c}i\v{c}}
\author{J\"org Main}
\author{G\"unter Wunner}
\affiliation{Institut f\"ur Theoretische Physik 1, Universit\"at Stuttgart,
  70550 Stuttgart, Germany}
\date{\today}

\begin{abstract}
In the preceding paper [T.\ Fab\v{c}i\v{c} et al., preprint]
``restricted Gaussian wave packets'' were introduced for the regularized 
Coulomb problem in the four-dimensional Kustaanheimo-Stiefel coordinates, 
and their exact time propagation was derived analytically in a 
fictitious time variable.  We now establish the Gaussian wave packet method 
for the hydrogen atom in static external fields.  A superposition of 
restricted Gaussian wave packets is used as a trial function in the 
application of the time-dependent variational principle.  The external 
fields introduce couplings between the basis states.  The set of coupled 
wave packets is propagated numerically, and eigenvalues of the Schr\"odinger 
equation are obtained by the frequency analysis of the time autocorrelation 
function.  The advantage of the wave packet propagation in the fictitious 
time variable is that the computations are exact for the field-free hydrogen 
atom and approximations from the time-dependent variational principle only 
 stem from the external fields.  Examples are presented for the hydrogen atom 
in a magnetic field and in crossed electric and magnetic fields.
\end{abstract}


\maketitle

\section{Introduction}
\label{sec:intro}
The hydrogen atom in a static magnetic field 
\cite{Hol88,Fri89,Has89,Mai94b,Fab05}
and in crossed electric and magnetic fields 
\cite{Wie89,Mai92,Mai94,Mil96,Neu97,Fre02,Bar03a,Bar03b,Gek06,Gek07,Car07a} 
is a non-integrable system which can be accessed both experimentally and 
theoretically and has attracted much attention during recent decades.
Exact quantum spectra of the system can be obtained by numerical
diagonalization of the Hamiltonian in a large Sturmian type basis set.
Nevertheless, the atom has served as an example system for the development 
and verification of alternative quantization methods, e.g., semiclassical
closed-orbit theory \cite{Du88,Bog89}, 
periodic-orbit theory \cite{Gut90,Bra97},
and cycle-expansion techniques \cite{Tan96}.

Another alternative to large quantum computations is the application of
the time-dependent variational principle (TDVP) \cite{McL64}.
For a wave packet depending on a set of variational parameters the
time-dependent Schr\"odinger equation is transformed to a system of
ordinary differential equations for the variational parameters.
Quantum spectra can be obtained by a frequency analysis of the time
autocorrelation function of the wave packet.
The method has been established by Heller \cite{Hel75,Hel76}
for single or coupled Gaussian wave packets (GWPs).
It is well suited for nonsingular smooth potentials but certainly 
far from ideal for atomic systems with singular Coulomb potentials.

The wave packet dynamics in atomic systems has been studied for the
field-free hydrogen atom \cite{Barnes93,Barnes94,Barnes95}, and in 
particular for the atom in time-dependent external fields, e.g., 
microwaves or short laser pulses.
While Rydberg wave packets are usually dispersive, the possible existence 
of nondispersive coherent states has been demonstrated for the 
hydrogen atom in 
microwave fields \cite{Buch95,Cer97}.

In the preceding paper \cite{Fab08a} we have established the Gaussian wave 
packet method for the Coulomb problem.
Using the Kustaanheimo-Stiefel (KS) regularization the singular Coulomb 
problem was transformed to the four-dimensional (4D) harmonic oscillator 
with a constraint.
We introduced the set of ``restricted Gaussian wave packets'' obeying
that constraint by confining the space of the Gaussian parameters.
The exact propagation of the restricted GWPs in a fictitious time variable 
could be derived analytically.

In this paper we extend the fictitious time wave packet propagation to the 
hydrogen atom in static external electric and magnetic fields.
A superposition of restricted GWPs is used as the variational trial function.
The time-dependent variational principle is applied in such a way that
the wave packet dynamics is exact for the field-free hydrogen atom
and couplings between the GWPs are only induced by the external fields.
In the presence of a single external homogeneous field the rotational 
symmetry of the hydrogen atom is preserved and one component 
of the angular momentum, say $l_z$, is conserved.
In that case we employ the modified 2D Gaussian wave packets with well-defined 
magnetic quantum number $m$ introduced and discussed in Ref.\ \cite{Fab08a}
and perform computations in the subspaces of the different magnetic 
quantum numbers $m$ separately.
In crossed fields the cylindrical symmetry is broken and computations
are performed in the basis of the restricted 3D GWPs without well-defined 
angular momentum quantum numbers. 

The fact that the wave packet propagation is exact for the pure Coulomb
problem might imply that the external fields are treated as a perturbation
and the method does not work well beyond the perturbative regime.
However, this is not the case.
The dynamics of wave packets is exact to all orders in the field strengths
within the allowed set of trial wave functions, i.e., the variational 
approximation only concerns the restriction of the Hilbert space.
The power of the method will be demonstrated by application to the 
diamagnetic hydrogen atom in the strong non-pertubative regime at the 
field-free ionization threshold.

The paper is organized as follows.
In Sec.\ \ref{sec_reg_extern} we introduce the regularization and scaling 
of the Hamiltonian with external fields and discuss the general idea of 
how to obtain quantum spectra by frequency analysis of the fictitious time 
autocorrelation function of the propagated wave packets.
In Sec.\ \ref{sec:TDVP} the time-dependent variational principle is explained.
The equations of motion for the variational parameters are derived for 
the superposition of restricted 3D and modified 2D GWPs, and the numerical time 
propagation of coupled wave packets is discussed.
Results for the diamagnetic hydrogen atom and the atom in crossed electric 
and magnetic fields are presented in Sec.\ \ref{sec:results}.
Concluding remarks are given in Sec.~\ref{sec:conclusion}.

\section{Regularized hydrogen atom in external fields}
\label{sec_reg_extern}
In the preceding paper \cite{Fab08a} the fictitious time wave packet dynamics
has been discussed for the field-free hydrogen atom.
We now consider the atom in external electric and magnetic fields.
For perpendicular fields with the electric and magnetic field along the $x$
and $z$ axis, respectively, 
 the Hamiltonian in the three-dimensional coordinates reads
(in atomic units with $F_0=5.14\times 10^9$V/cm, $B_0=2.35\times 10^5$T)
\begin{equation}
 H_3 = \frac{1}{2}{\bf p}^2 - \frac{1}{r} + \frac{1}{2}Bl_z
     + \frac{1}{8}B^2(x^2+y^2) + Fx \; .
\end{equation}
The starting point for our investigations is the Schr\"odinger equation 
in the 4D Kustaanheimo-Stiefel coordinates ${\bf u}$ with 
$x=u_1u_3-u_2u_4$, $y=u_1u_4+u_2u_3$, and 
$z=\frac{1}{2}(u_1^2+u_2^2-u_3^2-u_4^2)$.
Introducing scaled coordinates and momenta
${\bf u}\to n_{\rm eff}^{1/2}{\bf u}$, ${\bf p}_u\to n_{\rm eff}^{-1/2}{\bf p}_u$
and following the procedure of Sec.~II in \cite{Fab08a} we obtain
\begin{eqnarray}
     H\psi
 &=& \biggl\{ \frac{1}{2}{\bf p}_u^2 
      + \left[-n_{\rm eff}^2E
      + \frac{1}{8}(n_{\rm eff}^2B)^2(u_1^2+u_2^2)(u_3^2+u_4^2)  \right. \nonumber \\
  &&  \left.  + n_{\rm eff}^3F(u_1u_3-u_2u_4)\right]{\bf u}^2 \nonumber \\
 &+& \frac{1}{2}n_{\rm eff}^2B \left[(u_1p_2-u_2p_1)(u_3^2+u_4^2) \right. \nonumber \\
 && \left.    +(u_3p_4-u_4p_3)
     (u_1^2+u_2^2)\right] \biggr\}\psi = 2n_{\rm eff}\, \psi \; .
\label{regH4D}
\end{eqnarray}
In KS coordinates physical wave functions must fulfill the constraint
\begin{equation}
 (u_2 p_1 - u_1 p_2 - u_4 p_3 + u_3 p_4)\, \psi = 0 \; .
\label{eq_rest}
\end{equation}
By choosing constant parameters
\begin{equation}
 \alpha \equiv -n_{\rm eff}^2E\; ,\quad
 \beta \equiv n_{\rm eff}^2B \; ,\quad
 \zeta \equiv n_{\rm eff}^3F \; , 
\label{eq_neff_e_feld}
\end{equation}
Eq.\ \eqref{regH4D} becomes an eigenvalue problem for the effective quantum
number $n_{\rm eff}$.
For a set of parameters $(\alpha,\beta,\zeta)$ and a given 
eigenvalue $n_{\rm eff}$ the energy and field strengths of the physical 
state are obtained from Eq.\ \eqref{eq_neff_e_feld}.
The quantized energies and field strengths are located on lines 
 with constant $E/B$ and $E/F^{2/3}$.

In analogy with the field-free hydrogen atom in \cite{Fab08a} we can extend 
Eq.\ \eqref{regH4D} to the time-dependent Schr\"odinger equation in the
dimensionless fictitious time $\tau$ by the replacement 
$2n_{\rm eff}\to i\frac{\partial}{\partial\tau}$, viz.\
\begin{equation}
   i\frac{\partial}{\partial\tau}\psi
 = \left(\frac{1}{2}{\bf p}_u^2+V\right)\psi 
 = \left(-\frac{1}{2}\Delta_4+V\right)\psi
 = (T+V)\psi \; ,
\label{eq_TVpsi}
\end{equation}
where $V$ is defined via Eq.\ \eqref{regH4D} as the sum of a harmonic 
potential and the contributions of the external fields.
For the field-free hydrogen atom, i.e., a harmonic potential $V$, wave
packets can be propagated analytically in the fictitious time \cite{Fab08a}.

Our goal for the hydrogen atom in external fields is to compute the 
propagation of an initial wave packet $\psi(0)$ by applying the 
time-dependent variational principle.
To this end the wave function is assumed to depend on a set of appropriately chosen
parameters whose time-dependence is obtained by solving ordinary differential 
equations.
The ansatz for the wave function depends on the symmetry of the problem.
For the hydrogen atom in crossed fields we choose a superposition of $N$
restricted Gaussian wave packets \cite{Fab08a}
\begin{equation}
 \psi(\tau) = \sum_{k=1}^N e^{i[{\bf u}A_k(\tau){\bf u}+\gamma_k(\tau)]} \; ,
\label{eq_psi_tau}
\end{equation}
with the symmetric width matrices
\begin{equation}
 A = \left(\begin{array}{crcr}
  a_{\mu} & 0\; & a_{x} & a_{y} \\
  0      & a_{\mu} & a_{y} & -a_{x} \\
  a_{x} & a_{y} & a_{\nu} & 0\; \\
  a_{y} & -a_{x} & 0     & a_{\nu} 
\end{array}\right) \; ,
\label{eq:A_sym}
\end{equation}
depending on the four parameters $(a_\mu,a_\nu,a_x,a_y)$, and with $\gamma$ 
determining the normalization and phase of the restricted GWPs.
The special form of the ansatz \eqref{eq_psi_tau}, which depends on, in total,
 $5N$ 
time-dependent variational parameters (instead of $15N$ complex 
parameters for the
most general superposition of Gaussian wave packets in a 4D 
coordinate space) guarantees that the wave function obeys the constraint 
\eqref{eq_rest}.

The hydrogen atom in a pure magnetic field, i.e., $\zeta=0$, is cylindrically
symmetric around the $z$ axis, and the angular momentum component $l_z=m$ is
an exact quantum number.
For wave packets with given $m$ quantum number we use the ansatz
\begin{eqnarray}
   \psi_m(\tau)
& = & (\mu\nu)^{|m|} \sum_{k=1}^N
   e^{i\left[{\bf u}A_k(\tau){\bf u}+\gamma_k(\tau)\right]} e^{im\varphi} \nonumber \\
&  = &(\mu\nu)^{|m|} \sum_{k=1}^N
   e^{i\left[a_\mu^k(\tau)\mu^2+a_\nu^k(\tau)\nu^2+\gamma_k(\tau)\right]}e^{im\varphi} ,
\label{eq_psi_tau_m}
\end{eqnarray}
with the diagonal form of the matrix $A$ obtained by setting $a_x=a_y=0$ 
in \eqref{eq:A_sym}, and semiparabolic coordinates 
$\mu=\sqrt{u_1^2+u_2^2}=\sqrt{r+z}$ and $\nu=\sqrt{u_3^2+u_4^2}=\sqrt{r-z}$ 
are introduced. 
The wave function \eqref{eq_psi_tau_m} thus depends on a set of $3N$ 
time-dependent variational parameters.
As the paramagnetic term is constant this term can be absorbed by an energy
shift $E\to E'=E-mB/2$.
In semiparabolic coordinates the kinetic and potential term in \eqref{eq_TVpsi}
for the diamagnetic hydrogen atom then take the form
\begin{eqnarray}
 T &=& -\frac{1}{2}\left(\frac{\pa^2}{\pa \mu^2}
   + \frac{1}{\mu}\frac{\pa}{\pa \mu} - \frac{m^2}{\mu^2}
   + \frac{\pa^2}{\pa \nu^2}+ \frac{1}{\nu}\frac{\pa}{\pa \nu}
   - \frac{m^2}{\nu^2} \right) \; , \nonumber  \\
\label{4dHsemiV}
 V &=& \alpha(\mu^2+\nu^2) + \frac{1}{8}\beta^2(\mu^4\nu^2+\mu^2\nu^4) \; .
\end{eqnarray}

Once the time-dependent wave packets \eqref{eq_psi_tau} or \eqref{eq_psi_tau_m}
are determined the eigenvalues $n_{\rm eff}^{(j)}$ of the stationary 
Schr\"odinger equation \eqref{regH4D} and thus quantum spectra of the 
hydrogen atom in external fields are obtained by a frequency analysis of 
the time signal
\begin{equation}
 C(\tau) = \langle\psi(0)|\psi(\tau)\rangle 
 = \sum_j c_j e^{-i2n_{\rm eff}^{(j)}\tau} \; ,
\label{eq:C_tau}
\end{equation}
with the amplitudes $c_j$ depending on the choice of the initial wave packet.
The advantage of using the fictitious time $\tau$ is that the computations 
are exact for the field-free hydrogen atom and approximations from the
time-dependent variational principle only stem from the external fields.
By contrast, wave packet propagation in the physical time $t$ is a very
nontrivial task even for the pure Coulomb potential.

\section{Time-dependent variational principle}
\label{sec:TDVP}
\begin{figure}
\includegraphics[width=0.5\columnwidth]{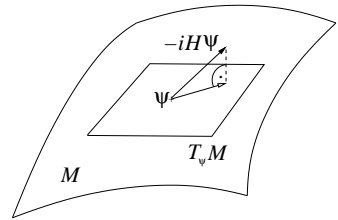}
\caption{Sketch of the manifold $M$ of approximation of the trial wave
  function $\chi(\z)$.  The variational evolution of the trial function,
  denoted by the arrow with the white arrowhead, is obtained as the projection of
  the exact time evolution $-iH\chi$, denoted by the arrow with the black arrowhead,
  onto the tangent space $T_\chi M$ of the manifold $M$ in the point $\chi$.}
\label{hilbert_skizze}
\end{figure}

The propagation of the wave packets investigated in this paper is based on 
the application of the time-dependent variational principle.
For the convenience of the reader we first give a brief general introduction
to the TDVP which is then applied to the special form of the trial functions
\eqref{eq_psi_tau} and \eqref{eq_psi_tau_m}.
The formulation of McLachlan \cite{McL64}, or equivalently the minimum error 
method \cite{Saw85}, requires the norm of the deviation between
the right-hand and the left-hand side of the time-dependent Schr\"odinger equation 
 to be minimized with respect to the trial function.
The quantity 
\begin{equation}
 I = ||i \phi(t) -H \chi(t)||^2 \overset{!}{=} \min
\label{mem}
\end{equation}
is to be varied with respect to $\phi$ only, and then $\dot\chi\equiv \phi$ 
is chosen, i.e., for any time $t$ the fixed wave function $\chi(t)$ is supposed 
to be given and its time derivative $\dot \chi(t)$ is determined by the 
requirement to minimize $I$. The equality 
 $I=0$ is provided by the exact solution of the Schr\"odinger equation, while
$I$ in general takes positive values if $\dot \chi $ is 
constrained by the functional form of $\chi$. 
The wave function $\chi(t)$ is assumed to be parametrized by a set of 
complex parameters $\z(t)=(z_1(t),\dots,z_{n_p}(t))$, $\chi(t)=\chi(\z(t))$.
For brevity the arguments of the wave function are dropped in the following.
For parametrized trial functions the variations $\delta \phi$ carry over to 
variations $\delta \dot \z$ and the variation leads to the equations of motion
\begin{equation}
 \lela \frac{\partial\chi}{\partial\z} \Big| i\dot\chi - H \chi \rira = 0 \; ,
\label{mem2}
\end{equation}
which can be written in matrix form
\begin{equation}
 K\dot \z = -i {\bf h} \quad \text{with} \quad
 K = \lela\frac{\partial \chi}{\partial \z}\Big|
          \frac{\partial \chi}{\partial \z}\rira \; , \; 
 {\bf h} = \lela \frac{\partial \chi}{\partial \z} \Big| H \chi \rira \; .
\label{mem3}
\end{equation}
An illustration of Eq.\ \eqref{mem2} is presented in 
Fig.\ \ref{hilbert_skizze}.
Here the manifold of approximation $M$, consisting of all possible 
configurations $\chi(\z)$, is plotted schematically as a $2$D-surface in the 
Hilbert space. 
The tangent space of the manifold in the point $\chi$ is a linear vector space 
and is spanned by the derivatives $\frac{\pa \chi}{\pa z_k},\,k=1,\dots,n_p$.
The tangent space is denoted by $T_\chi M$ in Fig.\ \ref{hilbert_skizze}.
According to the Schr\"odinger equation the exact time derivative $\dot \chi$ 
is given by $-iH\chi$, denoted by the arrow with the black arrowhead.
In general the exact time derivative does not lie in the tangent space,
otherwise the trial function would be an exact solution of the Schr\"odinger 
equation.
The variational approximation to the exact time derivative is given by that 
vector of the tangent space which has minimal deviation from the exact one.
This is the orthogonal projection of the exact time derivative onto the 
tangent space, denoted by the arrow with the white arrowhead in Fig.\ 
\ref{hilbert_skizze}.   

For parametrized wave functions the variational principle \eqref{mem} simply 
reduces to a quadratic minimization problem where the gradient of $I$ with 
respect to the time derivatives of the parameters must be zero 
\begin{equation}
 \frac{\partial I}{\partial\dot z_{k}} = 0 \; , \quad k=1,\dots,n_{p} \; ,
\end{equation}
and the TDVP leads to a reduction of the Schr\"odinger equation to a system 
of ordinary first-order differential equations of motion for the parameters 
$\z(t)$.
The matrix equation \eqref{mem3} must be solved numerically after each time 
step of integration for the time derivatives $\dot \z$ if a numerical 
algorithm for ordinary differential equations, e.g.\ Runge-Kutta or Adams, 
is used.

We now apply the time-dependent variational principle, first in 
Sec.\ \ref{sec_psi_m_var} to the trial function \eqref{eq_psi_tau_m}
of the diamagnetic hydrogen atom, and then in Sec.\ \ref{sec_psi_var} 
to the trial function \eqref{eq_psi_tau} of the hydrogen atom in 
crossed electric and magnetic fields.
For Gaussian type trial functions it is convenient to split the Hamiltonian
into the kinetic and potential part, i.e., $H=T+V$, and to apply Eq.\
\eqref{mem2} in the form
\begin{equation}
 \lela \frac{\partial\chi}{\partial\z} \Big| i\dot\chi - T \chi \rira
  =  \lela \frac{\partial\chi}{\partial\z} \Big| V \chi \rira \; .
\label{mem4}
\end{equation}
Note that the variational approach substantially differs from a perturbative
treatment of the hydrogen atom in external fields, and is valid even in the
strong non-perturbative regime.

\subsection{Diamagnetic hydrogen atom}
\label{sec_psi_m_var}
For the time-dependent wave packets of the hydrogen atom in a homogeneous 
external magnetic field with given $m$ quantum number we use the ansatz 
\eqref{eq_psi_tau_m} which can be written in the form
\begin{equation}
 \psi_m = \psi_m(\z) = \sum_{k=1}^N g_m(\y^k) \; ,
\label{eq_psi_m}
\end{equation}
with the basis states
\begin{equation}
 g_m(\y)=(\mu \nu)^{|m|}e^{i(a_\mu\mu^2+a_\nu\nu^2+\g)} \; .
\label{eq_KS_gwp_x_lz_m}
\end{equation}
As already mentioned, the cylindrical symmetry of the system is 
accounted for by setting $a_x=a_y=0$
and only the time-dependent parameters $\z=(\y^1,\dots,\y^N)$ with 
$\y=(\g,a_\mu,a_\nu)$ remain.
The evolution of the basis states is obtained by the TDVP.
The variational equations of motion are set up by evaluating Eq.\ \eqref{mem4}.
First we let the time derivative and the Laplacian act on the basis
states \eqref{eq_KS_gwp_x_lz_m} to obtain
\begin{eqnarray}
 & & \left( i \frac{\pa}{\pa \tau}-T \right) g_m({\bf y}^k,\x)  
  \nonumber \\
 &=& \left[-\dot \gamma^k+2i\left(a_\mu^k+a_\nu^k \right) 
 \left( 1 + |m| \right) - \right. \nonumber \\
& & \left. 
(\dot a_\mu^k + 2(a_\mu^k)^2) \mu^2
  -(\dot a_\nu^k+2 (a_\nu^k)^2) \nu^2  \right] g_m({\bf y}^k,\x) \nonumber \\
 &\equiv & \left[ v_0^k +
 \frac{1}{2}\left(V_\mu^k\mu^2+V_\nu^k \nu^2\right)\right]g_m({\bf y}^k,\x) \; ,
\label{eq_psi_dot_H_psi}
\end{eqnarray}
for $k=1,\ldots,N$.
Eq.\ \eqref{eq_psi_dot_H_psi} defines the coefficients $v_0^k$, $V_\mu^k$, 
$V_\nu^k$ as functions of the parameters $a_\mu^k$, $a_\nu^k$ and
the time derivatives $\dot \g^k$, $\dot a_\mu^k$, $\dot a_\nu^k$.
The equations of motion can be written as
\begin{subequations}
\label{eq_dot_lz_m_comp_1}
\begin{align}
 \dot a_\mu^k & =  -2(a_\mu^k)^2-\frac{1}{2} V_\mu^k \; , \label{eq_blbl} \\
 \dot a_\nu^k & =  -2(a_\nu^k)^2-\frac{1}{2} V_\nu^k \; , \label{eq_blbl1}\\
 \dot \gamma^k & = 2i\left(a_\mu^k + a_\nu^k \right)\left(1+|m|\right)-v_0^k \; ,
\end{align}
\end{subequations} 
with $k=1,\dots,N$, and the yet unknown coefficients $V_\mu^k$, $V_\nu^k$, 
and $v_0^k$.
Note that the equations of motion \eqref{eq_dot_lz_m_comp_1} are in general
coupled through the coefficients $v_0^k$, $V_\mu^k$, $V_\nu^k$ which become
time-dependent in the presence of anharmonic potentials.
They must be determined from a system of linear equations, which follows 
from Eq.\ \eqref{mem4} when inserting the trial function \eqref{eq_psi_m}.
Using the derivatives of the basis states \eqref{eq_KS_gwp_x_lz_m} 
with respect to the variational parameters, viz.\
$\frac{\pa}{\pa\gamma^k} g^k_m = ig^k_m$,
$\frac{\pa}{\pa a_\mu^k} g^k_m = i\mu^2g^k_m$, and
$\frac{\pa}{\pa a_\nu^k} g^k_m = i\nu^2g^k_m$,
Eq.\ \eqref{mem4} of the TDVP finally yields the matrix equation 
\begin{widetext}
\vskip -5ex
\begin{eqnarray}
     \sum_{k=1}^N \left( \langle g^l_m|g^k_m\rangle v_0^k 
     + \frac{1}{2}\langle g^l_m|\mu^2|g^k_m\rangle V_\mu^k
     + \frac{1}{2}\langle g^l_m|\nu^2|g^k_m\rangle V_\nu^k \right)
 &=& \sum_{k=1}^N \langle g^l_m|V(\mu,\nu)|g^k_m\rangle \; , \nonumber \\
     \sum_{k=1}^N \left( \langle g^l_m|\mu^2|g^k_m\rangle v_0^k
     + \frac{1}{2}\langle g^l_m|\mu^4|g^k_m\rangle V_\mu^k
     + \frac{1}{2}\langle g^l_m|\mu^2\nu^2|g^k_m\rangle V_\nu^k \right)
 &=& \sum_{k=1}^N \langle g^l_m|\mu^2V(\mu,\nu)|g^k_m\rangle \; , \nonumber \\
     \sum_{k=1}^N \left( \langle g^l_m|\nu^2|g^k_m\rangle v_0^k
     + \frac{1}{2}\langle g^l_m|\mu^2\nu^2|g^k_m\rangle V_\mu^k
     + \frac{1}{2}\langle g^l_m|\nu^4|g^k_m\rangle V_\nu^k \right)
 &=& \sum_{k=1}^N \langle g^l_m|\nu^2V(\mu,\nu)|g^k_m\rangle \; ,
\label{eq_linGl_l_m}
\end{eqnarray}
\end{widetext}
where the index $l=1,\dots ,N$ runs over all basis states and the notation 
$g^k_m \equiv g_m({\bf y}^k)$ is used.
The potential $V(\mu,\nu)$ for the diamagnetic hydrogen atom is given in
Eq.\ \eqref{4dHsemiV}.
All integrals in Eq.\ \eqref{eq_linGl_l_m} can be obtained analytically, 
and are presented in Appendix \ref{app_integr_dia}.
The set of equations \eqref{eq_linGl_l_m} is a $3N$-dimensional Hermitian 
positive semidefinite linear system for the coefficients $v_0^k$, $V_\mu^k$, 
$V_\nu^k$,  $k=1,\dots,N$, and must be solved, e.g., using a Cholesky 
decomposition \cite{nr} of the left-hand side matrix, at every time step when 
numerically integrating the equations of motion \eqref{eq_dot_lz_m_comp_1}.
Technical remarks for the time propagation of coupled wave packets via
the numerical integration of the Eqs.\ \eqref{eq_dot_lz_m_comp_1} will
be given in Sec.\ \ref{sec:num_TDVP}.

\subsection{Hydrogen atom in crossed fields}
\label{sec_psi_var}
The rotational symmetry of the hydrogen atom in a magnetic field as discussed 
in Sec.\ \ref{sec_psi_m_var} is broken when an additional electric field 
with a different orientation is applied.
In crossed fields none of the three degrees of freedom can be separated.
The paramagnetic term that contributed only a constant energy shift within 
the subspace of constant $m$ in the diamagnetic hydrogen atom must now be 
taken into account since $l_z$ is not conserved.
The evolution of wave packets is determined by the time-dependent 
Schr\"odinger equation \eqref{eq_TVpsi} with $T$ given by minus one half times
the Laplacian in the 4D Kustaanheimo-Stiefel coordinates,
and $V$ defined via Eqs.\ \eqref{regH4D} and \eqref{eq_neff_e_feld} as
\begin{eqnarray}
 V &=& \alpha{\bf u}^2 + \frac{1}{2}\beta
       [(u_1p_2-u_2p_1)(u_3^2+u_4^2) \nonumber \\
   && +(u_3p_4-u_4p_3)(u_1^2+u_2^2)]  \\
   &+& \frac{1}{8}\beta^2(u_1^2+u_2^2)(u_3^2+u_4^2){\bf u}^2
       + \zeta (u_1u_3-u_2u_4){\bf u}^2 \; .\nonumber 
\label{eq:V}
\end{eqnarray}
As trial functions for the time-dependent variational principle we use
the superposition
\begin{equation}
 \psi(\z) = \sum_{k=1}^N g(\y^k)
\label{eq_psi}
\end{equation}
where 
\begin{equation}
 g^k \equiv g(\y^k) = e^{i(\uu A^k \uu + \gamma^k)}
\label{eq:g_k}
\end{equation}
are the restricted Gaussian wave packets derived in the preceding paper 
\cite{Fab08a}, which depend on the $5N$ time-dependent variational parameters
$\y^k=(\gamma^k,a_\mu^k,a_\nu^k,a_x^k,a_y^k)$ (see Eq.\ \eqref{eq:A_sym}), combined 
in the parameter vector $\z=(\y^1,\dots,\y^N)$. 
The equations of motion for the variational parameters are obtained by
evaluating the TDVP in Eq.\ \eqref{mem4} for the trial function \eqref{eq_psi}.
The procedure is similar to that in Sec.\ \ref{sec_psi_m_var}.
Letting the time derivative and the Laplacian act on a
restricted GWP \eqref{eq:g_k} yields
\begin{eqnarray}
       \left(i\frac{\pa}{\pa \tau}-T\right) g^k
 & = & \left(-\uu\dot A^k\uu-\dot \gamma^k 
       - 2\uu(A^k)^2\uu  \right. \nonumber \\
 {}+ i\,{\rm tr}\,A^k\Big) g^k 
 &\equiv& \left(v_0^k+\frac{1}{2}\uu V_2^k\uu \right) g^k \; ,
\label{v2_def}
\end{eqnarray}
and defines a scalar $v_0^k$ and a $4\times 4$ matrix $V_2^k$ as the 
coefficients of the polynomial in $\uu$ for each GWP with $k=1,\dots,N$, 
i.e., $v_0^k=i\,{\rm tr}\, A^k-\dot\gamma^k$ and $V_2^k/2=-\dot A^k-2(A^k)^2$.
Since the special structure of the matrices $A^k$ in Eq.\ \eqref{eq:A_sym} is 
maintained in the squared matrices $(A^k)^2$, that structure carries over to
the $ 4\times 4$ complex symmetric matrices $V_2^k$ due to their definition
in Eq.\ \eqref{v2_def}. 
Therefore, they have only four independent coefficients $V_\mu^k$, $V_\nu^k$,
$V_x^k$, and $V_y^k$ in the notation of Eq.\ \eqref{eq:A_sym}.
The equations of motion for the variational parameters
$\y^k=(\gamma^k,a_\mu^k,a_\nu^k,a_x^k,a_y^k)$, $k=1,\dots,N$ 
can be written as
\begin{subequations}
\label{eq_gwp_reg_eq_motion_var_both}
\begin{align}
\label{eq_gwp_reg_eq_motion_var}
 \dot A^k &=  -2(A^k)^2-\frac{1}{2}V_2^k \; , \\
 \dot \g^k &= i\, {\rm tr}\,A^k -v_0^k \; ,
\end{align}
\end{subequations}
where the time-dependent parameters $(v_0^k,V_\mu^k,V_\nu^k,V_x^k,V_y^k)$ 
are obtained at every time step by solving a linear set of equations.
Using the derivatives of the restricted GWPs with respect to the variational 
parameters,
\begin{eqnarray}
 \frac{\pa g^k}{\pa\g^k} &=& ig^k \; , \quad 
 \frac{\pa g^k}{\pa a_\mu^k}  =  i(u_1^2+u_2^2)g^k \; , \quad \nonumber \\
 \frac{\pa g^k}{\pa a_\nu^k}  & = &  i(u_3^2+u_4^2)g^k \; ,\quad  
 \frac{\pa g^k}{\pa a_x^k}  = 2i(u_1u_3-u_2u_4)g^k \; , \quad \nonumber \\
 \frac{\pa g^k}{\pa a_y^k} & = & 2i(u_1u_4+u_2u_3)g^k \; , 
\label{eq:g_deriv}
\end{eqnarray}
the required linear set of equations is derived from \eqref{mem4} as
\begin{widetext}
\vskip -5ex
\begin{eqnarray}
     \sum_{k=1}^N\left( I_{11}^{lk}v_0^k + I_{12}^{lk}\frac{1}{2}V_\mu^k
     + I_{13}^{lk}\frac{1}{2}V_\nu^k + I_{14}^{lk}V_x^k + I_{15}^{lk}V_y^k \right)
 &=& \sum_{k=1}^N I_{v1}^{lk} \; , \nonumber \\
     \sum_{k=1}^N\left( I_{12}^{lk}v_0^k + I_{22}^{lk}\frac{1}{2}V_\mu^k
      + I_{23}^{lk}\frac{1}{2}V_\nu^k + I_{24}^{lk}V_x^k + I_{25}^{lk}V_y^k \right)
 &=& \sum_{k=1}^N I_{v2}^{lk} \; , \nonumber \\
     \sum_{k=1}^N\left( I_{13}^{lk}v_0^k + I_{23}^{lk}\frac{1}{2}V_\mu^k
     + I_{33}^{lk}\frac{1}{2}V_\nu^k + I_{34}^{lk}V_x^k + I_{35}^{lk}V_y^k \right)
 &=& \sum_{k=1}^N I_{v3}^{lk} \; , \nonumber \\
     \sum_{k=1}^N\left( I_{14}^{lk}v_0^k + I_{24}^{lk}\frac{1}{2}V_\mu^k
     + I_{34}^{lk}\frac{1}{2}V_\nu^k + I_{44}^{lk}V_x^k + I_{45}^{lk}V_y^k \right)
 &=& \sum_{k=1}^N I_{v4}^{lk} \; , \nonumber \\
     \sum_{k=1}^N\left( I_{15}^{lk}v_0^k + I_{25}^{lk}\frac{1}{2}V_\mu^k
     + I_{35}^{lk}\frac{1}{2}V_\nu^k + I_{45}^{lk}V_x^k + I_{55}^{lk}V_y^k \right)
 &=& \sum_{k=1}^N I_{v5}^{lk} \; ,
\label{eq_lin_Gl_3D}
\end{eqnarray}
\end{widetext}
with $l=1,\dots,N$.
All integrals $I$ in Eq.\ \eqref{eq_lin_Gl_3D} are defined and listed in 
Appendix \ref{app_int_3D}.
The potential \eqref{eq:V} for the hydrogen atom in crossed electric and
magnetic fields, including the paramagnetic contribution, enters the 
integrals on the right-hand side of Eq.\ \eqref{eq_lin_Gl_3D}.
The linear set of equations \eqref{eq_lin_Gl_3D} is Hermitian positive 
semidefinite [see Eq.\ \eqref{eq_linGl_l_m} for the diamagnetic hydrogen atom]
and can be solved using a Cholesky decomposition of the left-hand side matrix.

\subsection{Numerical time propagation of coupled wave packets}
\label{sec:num_TDVP}
An initial wave packet given as the superposition of basis states in
Eq.\ \eqref{eq_psi_m} or \eqref{eq_psi} can be easily propagated for the 
field-free hydrogen atom because the basis states remain uncoupled and the
time-dependence of the basis states is known analytically \cite{Fab08a}.
The external fields lead to couplings between the basis states, and the
time-dependence of the variational parameters must be determined numerically.
The setup of the equations of motion has been discussed in Secs.\
\ref{sec_psi_m_var} and \ref{sec_psi_var}.  The numerical integration, 
however,  
of Eqs.\ \eqref{eq_dot_lz_m_comp_1} and \eqref{eq_gwp_reg_eq_motion_var_both}
is nontrivial and further remarks are necessary.

\subsubsection{Time propagation of the width matrices}
For better numerical performance 
it is advantageous \cite{Hel76a,Saw85} 
to introduce, for each width matrix $A$,  
two auxiliary time-dependent $4\times 4$ matrices $B$ and $C$ in such  a way 
that 
\begin{equation}
 A = \frac{1}{2} B C^{-1} \; .
\end{equation}
The equations of motion \eqref{eq_gwp_reg_eq_motion_var} and similarly
Eqs.\ \eqref{eq_blbl} and \eqref{eq_blbl1} are then replaced with the
equivalent differential equations
\begin{eqnarray}
 \dot B^k &=& -V_2^kC^k \; , \nonumber \\
 \dot C^k &=&  B^k \; ,
\label{eq_BC_motion}
\end{eqnarray}
with the initial values $B(0)=2A(0)$ and $C(0)={\boldsymbol 1}$.
In the case of the diamagnetic hydrogen atom the matrices $A$ and $V_2$ are
diagonal with diagonal elements $\{a_\mu,a_\mu,a_\nu,a_\nu\}$ and
$\{V_\mu,V_\mu,V_\nu,V_\nu\}$, respectively.
The matrices $B$ and $C$ have the same structure, and thus the total number 
of parameters per basis state that must be integrated (including the scalar
$\gamma$) increases from three parameters $(\gamma,a_\mu,a_\nu)$ to five 
parameters $(\gamma,b_\mu,b_\nu,c_\mu,c_\nu)$.

For crossed fields the increase of the number of parameters is even more rapid.
In that case the matrices $B$ and $C$ are no more complex symmetric.
Without taking care of the special structure \eqref{eq:A_sym} of the matrix $A$ 
the introduction of the $B$ and $C$ matrices would require the integration
of 32 complex parameters per GWP in the two matrices $B$ and $C$ instead of
four complex parameters in the width matrix $A$.
However, the special structure of the matrix $A$ can be exploited to halve the
number of independent parameters from 32 to 16 in the matrices $B$ and $C$.
Details are given in Appendix \ref{app_param_red}. 

When integrating the equations of motion most of the computational effort
is invested in solving the set of $3N$ linear equations \eqref{eq_linGl_l_m}
or the $5N$ linear equations \eqref{eq_lin_Gl_3D} at each time step.
The dimension of those equations is not affected by the introduction of
the auxiliary matrices $B$ and $C$, and thus the increase of the number
of parameters in the differential equations \eqref{eq_BC_motion} does not 
imply a significant increase of the total computing effort. In fact, 
due to the better numerical behavior of Eq.\ \eqref{eq_BC_motion} as compared
to Eq.\ \eqref{eq_gwp_reg_eq_motion_var} and 
Eqs.\ \eqref{eq_blbl}, \eqref{eq_blbl1}, larger step sizes of the numerical
integration are possible and the total computing time is decreased.

\subsubsection{TDVP with constraints}
\label{sec:constraints}
The equations of motion resulting from the TDVP especially for a large number 
of coupled GWPs become badly behaved from time to time during the integration.
In the general formulation of the TDVP at each time step the linear set of 
equations \eqref{mem3} must be solved for the equations of motion of the 
variational parameters, i.e., the time derivatives $\dot\z$.
In the course of integration, depending on the number of coupled GWPs,
it will happen sooner or later that the matrix $K$ in Eq.\ \eqref{mem3}
associated with the set of linear equations becomes ill-conditioned, 
or even numerically singular.
As a result the time step of the integration routine becomes extremely
small, rendering the method of GWP propagation impracticably slow.
In the worst case the wave packet propagation can stick completely.
 
Matrix singularity problems arise from overcrowding the basis set, i.e., from 
situations where fewer GWPs would be sufficient to represent the wave function.
On the other hand for an accurate approximation of the wave function it is 
desirable to have a large number of adjustable parameters.
However, there is a discrepancy between the number of GWPs necessary to yield 
accurate results and the maximum number of GWPs that can be propagated using 
the TDVP without numerical difficulties \cite{Han89}.
There exist different proposals to overcome this numerical problem 
\cite{Kay89,Hea86,Hor04,Saw85,Han89,Sko84,Zop05,Fab08}.
Here we adopt the constrained time-dependent variational principle 
\cite{Fab08}, where inequality constraints of the form
%
\begin{eqnarray}
  f_k(\z,\z^*) & \equiv & f_k(\z_r,\z_i) \equiv  f_k( \bar \z)
  \geq f_{k,\min} \; ,\nonumber \\
  \;f_k  & \in &  {\mathbb R},\; k=1,2,3,\ldots
  \label{const}
\end{eqnarray}
%
are taken into account in the variational process, and complex quantities 
are split into their real and imaginary parts, denoted by the subscripts 
$r$ and $i$, respectively, and thus $\bar\z \equiv (\z_r,\z_i)$.
 The functions $f_k$ must be chosen in such a way to prevent the matrix 
$K$ from becoming singular.
As long as $f_k(\z_r,\z_i)>f_{k,\rm min} $ for all $k$, all parameters evolve 
 according to 
Eq.\ \eqref{mem3} without being affected by the constraints.
However, when $f_k(\z_r,\z_i) = f_{k,\rm min}$ and $\dot f_k(\z_r,\z_i) <0 $ 
for, say, $k=1,\ldots,j$ 
we introduce Lagrangian multipliers and obtain an extended set of linear 
equations  
\begin{gather}
 \left(\begin{array}{c|c}
 \bar K & \bar M^T \\ \hline \bar M  & 0 \end{array} \right)
 \left(\begin{array}{c} \dot{\bar \z} \\ {\boldsymbol \lambda}
 \end{array} \right) 
 = \left( \begin{array}{c} \bar{\bf h} \\ 0 \end{array} \right) , \nonumber \\ \;
 {\rm with} \; \bar K =
 \left(\begin{array}{cr} K_r & -K_i \\ K_i &  K_r \\ \end{array} \right) , 
 \bar{\bf h} = \left(
 \begin{array}{r} {\bf h}_i \\  -{\bf h}_r \end{array} \right) \; ,
\label{ggcon}
\end{gather}
where the matrix $K$ and the vector ${\bf h}$ are the complex quantities of
Eq.\ \eqref{mem3}.
The Lagrangian multipliers are ${\boldsymbol \lambda} \in {\mathbb R}^j$ and
$\bar M = \frac{\partial {\bf f}}{\partial \bar \z}$ with ${\bf f}=(f_1,\ldots,f_j)$ 
is a real valued $j \times 2n_p$ matrix.
Details of the implementation of the constrained TDVP are given in \cite{Fab08}.
If no constraint is active, i.e., $j=0$, then Eq.\ \eqref{ggcon} obviously 
reduces to the real formulation of Eq.\ \eqref{mem3}.

\section{Results and discussion}
\label{sec:results}
In this section we present examples for the fictitious time wave packet 
propagation of the hydrogen atom in external fields.
Autocorrelation functions between the initial and time propagated wave
packets are computed.
Quantum spectra are obtained by the frequency analysis of the autocorrelation
function and compared with numerically exact diagonalizations of the
Hamiltonian.

It turns out that a sensible choice of an appropriate initial state 
$\psi(0)$ is crucial for the successful application of the TDVP.
For an  unreasonable choice the numerical problems discussed in 
Sec.\ \ref{sec:constraints} occur for few basis states already, and bad, 
unconverged results are obtained.
The conventional way to construct an initial wave packet by placing 
a certain number of unrestricted GWPs at various positions in coordinate 
and momentum space is not possible for the restricted GWPs in the KS 
coordinates.
In the calculations of the diamagnetic and the crossed fields hydrogen atom 
we achieved optimal results by first choosing only one 2D or 3D Gaussian wave 
packet in the physical coordinates,
which was then expanded in a set of $N$ restricted GWPs as explained in 
Ref.\ \cite{Fab08a} for the field-free hydrogen atom.
The external fields lead to couplings between the basis states and imply
a complicated time development of the initial state as compared to the 
field-free hydrogen atom, where the wave packet propagation is periodic 
in time \cite{Fab08a}.

\subsection{Diamagnetic hydrogen atom}
\label{sec:results_dia}
The initial wave function is most conveniently chosen to be a GWP in parabolic
coordinates 
\begin{equation}
   \psi(\xi,\eta)
 = A e^{-(\xi-\xi_0)^2/(4\sigma^2)-(\eta-\eta_0)^2/(4 \sigma^2)
     +ip_{\xi_0}(\xi-\xi_0)+ip_{\eta_0}(\eta-\eta_0)}, 
\label{eq_GWP_parabol}
\end{equation}
with center $(\xi_0,\eta_0)$, width $\sigma$ and mean momentum 
$(p_{\xi_0},p_{\eta_0})$.
The GWP is expanded in terms of the basis states \eqref{eq_KS_gwp_x_lz_m} 
according to the procedure described in detail in Ref.\ \cite{Fab08a}, 
including the Monte Carlo technique with importance sampling.
The procedure yields the initial values of the variational parameters 
$\g^k, a_\mu^k,a_\nu^k,\;k=1,\ldots,N$.

\begin{figure}
\includegraphics[width=0.98\columnwidth]{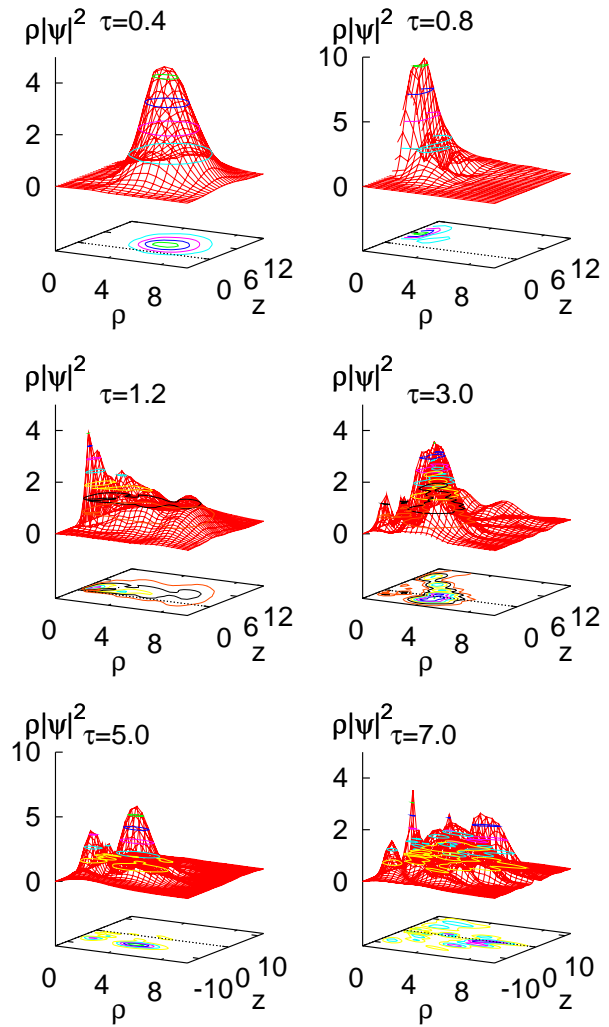}
\caption{(Color online) Fictitious time evolution of the state 
 \eqref{eq_GWP_parabol} with $\rho_0 = 6.0,z_0=0$ and a nonzero initial mean
 momentum.  The wave function is plotted for different values of the 
 dimensionless fictitious time $\tau$.  The initial wave packet gradually
 becomes delocalized.  Lengths are given in scaled atomic units
 $n_{\rm eff}a_0$ with $a_0$ the Bohr radius (see Eq.\ \eqref{regH4D}).}
\label{fig_wf_mult_m0_var}
\end{figure}
However, it is not realistic to propagate several thousands of basis states 
numerically with the full coupling.
Reliable results are obtained by far fewer basis states than used in the 
expansion and propagation of the GWP  \eqref{eq_GWP_parabol} in the 
field-free hydrogen atom \cite{Fab08a}. 
Reasonable numbers of basis states are in the range of N=10-100.
A numerical example is presented for the magnetic quantum number $m=0$, where 
 $N=70$ basis states are used for the expansion and propagation.
The damping factor $\epsilon$ 
is set to $\epsilon=0.1$

Each basis state has three variational parameters $\g^k,a_\mu^k,a_\nu^k$, 
and therefore $N$ basis states require the solution of a $3N \times 3N$ matrix
equation after every integration step, and the usual numerical problems 
mentioned in Sec.\ \ref{sec:TDVP} occur with increasing number of basis states.
It turns out that constraints on the imaginary parts of the phase parameters 
of the form ${\rm Im} \gamma^k \ge \gamma_{\rm min}=-4.5;\, k=1,\dots,N$ are
suitable to regularize the equations of motion with regard to a fast 
integration.
These constraints present simple lower bounds on the amplitudes of the wave 
packets and avoid matrix singularities caused by extremely large overlapping 
wave packets.

The accuracy of the expansion \eqref{eq_GWP_parabol} of the GWP with only 
$N=70$ basis states is very good. 
The time evolution of the wave function is shown in 
Fig.\ \ref{fig_wf_mult_m0_var}. 
The probability density $\rho|\psi(\rho,z)|^2$ for six different times 
$\tau = 0.4,0.8,1.2,3.0,5.0,7.0$ is shown. 
The parameters of the potential in the Hamiltonian \eqref{4dHsemiV} are set 
to $\alpha =0.5$ and $\beta = 0.2$.
The $\pi$ periodicity of the evolution of the wave function that is present 
in the field-free hydrogen atom, is destroyed now.

The autocorrelation function of the propagation
\begin{figure}
\includegraphics[width=0.95\columnwidth]{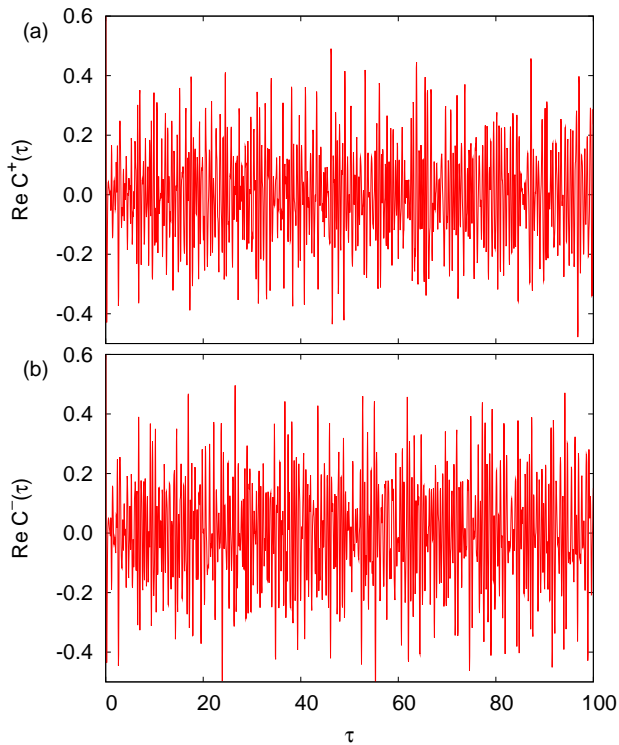}
\caption{(Color online) Real part of the autocorrelation function
 $C^\pm(\tau) = \langle \psi_0^\pm(0) | \psi_0^\pm(\tau) \rangle$ for the 
 GWP \eqref{eq_GWP_parabol} with the center $\rho_0 = 6.0,z_0=0$.
 (a) Signal of the projected state with even parity, and (b) odd parity.
 The fictitious time $\tau$ and the signal $C(\tau)$ are in dimensionless
 units.}
\label{fig_Ct_3D_70_dia}
\end{figure}
can be used to extract spectral information by Fourier transformation or 
harmonic inversion \cite{wall:8011,Man97,Mai99,Mai00,Bel00} of the time signal.
The center of the Gaussian \eqref{eq_GWP_parabol} is $\rho_0=6,z_0=0$ and the 
initial mean momentum is chosen in such a way that states around an effective 
quantum number of $n_{\rm eff} \approx 6$ are excited.

\begin{figure}
\includegraphics[width=0.95\columnwidth]{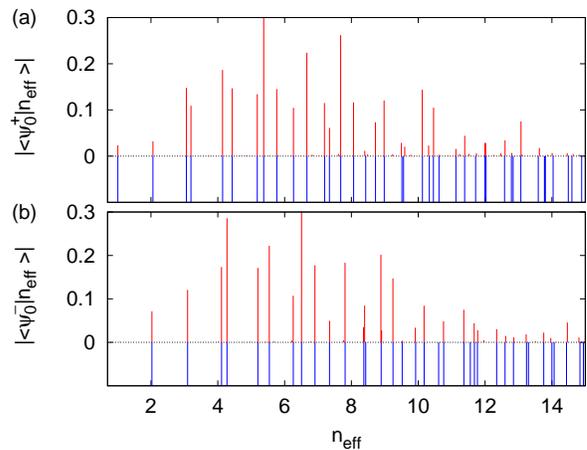}
\caption{(Color online) Spectra with (a) even and (b) odd $z$-parity extracted 
 from the autocorrelation function
 $C^\pm(\tau ) = \langle \psi_0^\pm(\tau=0)|\psi_0^\pm(\tau) \rangle $ 
 computed from the evolution of the wave function \eqref{eq_GWP_parabol} 
 plotted in Fig.\ \ref{fig_wf_mult_m0_var}.
 The amplitudes are given by the magnitude of overlap between the initial
 wave function and the respective eigenstates.  For comparison the positions
 of the numerically exact eigenvalues obtained from a diagonalization are
 plotted with blue lines in the lower panels of the figures.  The related
 eigenenergies and the magnetic field strength follow simply from
 Eq.\ \eqref{eq_neff_e_feld}.  The effective quantum number $n_{\rm eff}$
 and the overlap matrix elements are in dimensionless units.}
\label{fig_hi_m0_6_6_1_5}
\end{figure}

To reduce the density of states the autocorrelation function is separately 
computed for  the subspaces of even and odd parity by taking the symmetrized 
and antisymmetrized states 
$\psi^\pm_0(\rho,z) = \psi_0(\rho,z) \pm \psi_0(\rho,-z)$. 
The autocorrelation function 
$C^\pm(\tau) = \langle \psi_0^\pm(0) | \psi_0^\pm(\tau) \rangle$, is shown in 
Fig.\ \ref{fig_Ct_3D_70_dia}(a) for symmetrized states and in 
Fig.\ \ref{fig_Ct_3D_70_dia}(b) for the antisymmetric states.
The spectral results for the diamagnetic hydrogen atom, obtained from the 
time signals are plotted in Fig.\ \ref{fig_hi_m0_6_6_1_5}.
A harmonic inversion has been employed.
The amplitudes of the peaks are determined by the magnitude of the overlap 
between the eigenstates, denoted by $|n_{\rm eff} \rangle$, and the initial 
states $\psi^\pm_0(0)$ in Fig.\ \ref{fig_hi_m0_6_6_1_5}(a) and 
Fig.\  \ref{fig_hi_m0_6_6_1_5}(b), respectively.
The amplitudes are plotted with red lines.
The numerically exact eigenvalues of the diamagnetic hydrogen atom are plotted 
with blue lines for comparison. The agreement of the positions is excellent.
The highest amplitudes are located in the region $n_{\rm eff} \approx 6$ 
according to the choice of the input parameters of the initial GWP in 
Eq.\ \eqref{eq_GWP_parabol}.
The multiplicity of the states with even or odd $z$-parity resulting from 
the same principle quantum number $n$ is determined by the number of positive 
and negative eigenvalues $(-1)^{l+m}$ of the $z$-parity operator acting on 
the spherical harmonics $Y_{lm}(\theta,\varphi)$ with $l<n$.

\begin{figure}
\includegraphics[width=0.95\columnwidth]{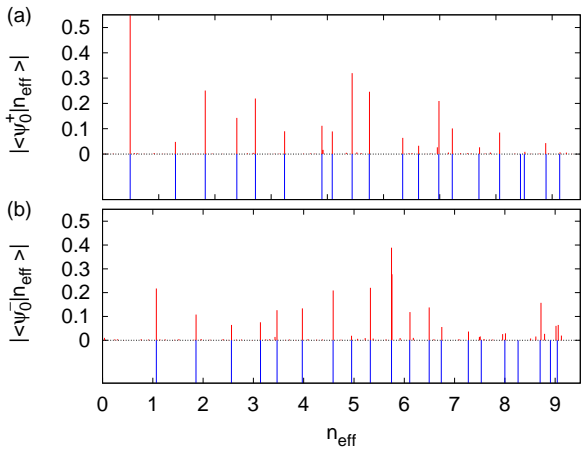}
\caption{(Color online) Effective quantum numbers $n_{\rm eff}$ at the
 field-free ionization threshold $E=\alpha=0$ for states of (a) even and (b)
 odd parity.  The propagation involves $N=90$ basis states.   The variational
 results (red lines in the upper panels) are in excellent agreement with the
 exact time-independent results (blue lines in the lower panels).
 The effective quantum number $n_{\rm eff}$ and the overlap matrix elements 
 are in dimensionless units.}
\label{fig_hiE0}
\end{figure}

The values of the parameters $\alpha = 0.5$ and $\beta=0.2$ used for this 
computation still present a mainly harmonic system with a perturbation for 
low energies.
As mentioned above the dynamics of wave packets is exact to all orders 
in the field strengths within the allowed set of trial wave functions, i.e., 
the variational approximation only concerns the restriction of the Hilbert 
space.
Therefore, the method is not restricted to the pertubative regime but even
allows for the computation of eigenvalues in the strong anharmonic regime.
Fig.\ \ref{fig_hiE0} presents results at the field-free ionization energy 
$E=\alpha =0$ and $\beta = 0.5$ for (a) even parity and (b) odd parity.
A number of $N=90$ basis states was used in the computation.     
In the presence of the magnetic field these states at the field-free 
ionization energy $E=0$ are still bound. 
The agreement between the eigenvalues computed variationally (red lines) and 
the numerically exact results (blue lines) is very good.
The related field strengths are easily obtained from Eq.\ \eqref{eq_neff_e_feld}
by $B=\beta/n_{\rm eff}^2$.
The underlying initial wave packet  \eqref{eq_GWP_parabol} is initially 
centered at $\rho_0=4.39,z_0=1$ and has zero mean momentum.  
As mentioned above the position, momentum, and width of the initial GWP
determine the spectral region for $n_{\rm eff}$ where strong peaks are
expected.
However, within that region some eigenstates $|n_{\rm eff}\rangle$
can be near orthogonal to the initial GWP and thus have nearly zero
amplitude.
Indeed, some lines are lacking in the variational computation.
The missing states can be revealed by choosing several initial GWPs,
which have larger overlap with those states.
An example of the influence of the chosen initial GWP on the peak amplitudes
will be given in Sec.\ \ref{sec:results_cross} for the hydrogen atom
in crossed electric and magnetic fields.

\subsection{Hydrogen atom in crossed fields}
\label{sec:results_cross}
For the hydrogen atom in crossed electric and magnetic fields the 
propagation of 3D GWPs is computed starting from the time-dependent
Schr\"odinger equation \eqref{regH4D} with parameters $\alpha = 0.5$, 
$\beta = 0.05$, and $\zeta = 0.01$ in Eq.\ \eqref{eq_neff_e_feld}.
The choice of an appropriate initial state $\psi(0)$ is very important
for the successful application of the TDVP.
We achieved optimal results by first choosing one 3D Gaussian wave packet 
in physical Cartesian coordinates with the center $\x_0$ and width $\sigma$ 
in position space and center $\p_0$ in momentum space
\begin{equation}
 \psi(\x) = (2 \pi \sigma^2)^{-3/4}
 \exp\left\{ -\frac{ (\x-\x_0)^2}{4 \sigma^2}+i\p_0\cdot (\x-\x_0) \right\}
\end{equation} 
which is then expanded in a set of $N$ restricted GWPs.
The external fields lead to couplings between the basis states, and the
time-dependence of the variational parameters must be determined by
the numerical integration of Eq.\ \eqref{eq_gwp_reg_eq_motion_var_both}.
For better numerical performance we resort to the TDVP with constraints 
\cite{Fab08} mentioned in Sec.\ \ref{sec:constraints}.
As for the diamagnetic hydrogen atom constraints of the form 
${\rm Im} \g^k \ge \gamma_{\rm min}=-4.0,\;k=1,\ldots,N$ are imposed on the 
imaginary parts of the phase parameters $\gamma^k$.
\begin{figure}
\includegraphics[width=0.95\columnwidth]{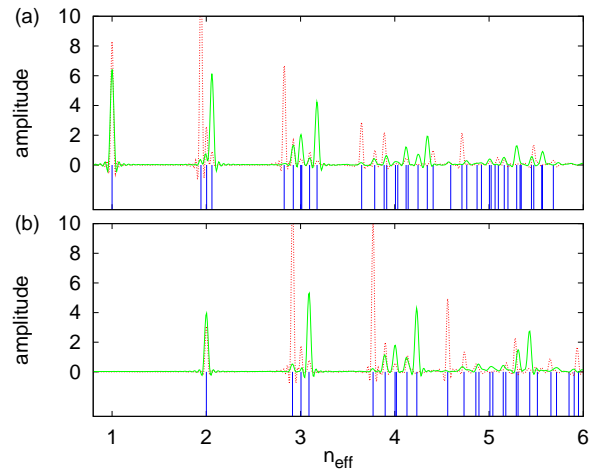}
\caption{(Color online) Spectra with (a) even and (b) odd $z$ parity of the
 Hamiltonian \eqref{regH4D} with $\alpha=0.5$, $\beta=0.05$, $\zeta = 0.01$
 obtained from the propagation of two different 3D GWPs.  Green and red line
 (upper panels in the figures): $\x_0=(6,0,0)$, $\p_{0}=(0,\pm 1/\sqrt{2},1/\sqrt{2})$,
 respectively.  The eigenvalues are extracted from the autocorrelation
 function by Fourier transformation.  The peak positions agree very well with
 the numerically exact eigenvalues of the effective quantum number
 marked by blue lines in the lower panels of the figures.  The related
 eigenenergies and the field strengths follow from Eq.\ \eqref{eq_neff_e_feld}.
 The effective quantum number $n_{\rm eff}$ and amplitudes are in dimensionless
 units.}
\label{fig_spek_cross_field}
\end{figure}
Once a time-dependent wave packet \eqref{eq_psi_tau} is determined the 
eigenvalues $n_{\rm eff}$ of the stationary Schr\"odinger equation 
\eqref{regH4D} are obtained by the frequency analysis of the time signal
\eqref{eq:C_tau} with the amplitudes $c_j$ depending on the choice of the 
initial wave packet.
In perpendicular crossed fields the $z$ parity is conserved.
Spectra with even and odd $z$ parity obtained from the Fourier transforms 
of the autocorrelation functions 
$C^{\pm}(\tau)=\langle\psi^{\pm}(0)|\psi^{\pm}(\tau)\rangle$ of the parity
projected wave packets are shown in Fig.\ \ref{fig_spek_cross_field}.
In Fig. \ref{fig_spek_cross_field}(a) the eigenvalues with even parity and in 
Fig. \ref{fig_spek_cross_field}(b) the eigenvalues with odd parity are plotted.
The green and red lines result from the propagation of two different 3D GWPs 
with $\sigma=3.5$, $\epsilon = 0.15$ and the same initial position 
$\x_0=(6,0,0)$  but different initial mean momenta 
$\p_{0}=(0,\pm 1/\sqrt{2},1/\sqrt{2})$, respectively.
A number of $N=41$ and $N=31$ basis states were coupled in the calculations.
The line widths, i.e., the resolution of the spectra, is determined by the
length of the time signal $\tau_{\max}$. 
The eigenvalues obtained by numerically exact diagonalizations of the
stationary Hamiltonian \eqref{regH4D} are shown by the blue lines.
The line-by-line comparison shows good agreement between the exact spectrum
and the results obtained by the wave packet propagation.  
The amplitudes of levels indicate the excitation strengths of states with
higher or lower angular momentum $l_z$ by the two initial wave packets
rotating clockwise or anticlockwise around the $z$ axis.

\section{Conclusion}
\label{sec:conclusion}
The Gaussian wave packet method is known to be well suited for systems with
nonsingular smooth potentials but not so for systems with singular potentials 
such as the Coulomb potential.
Therefore so far it failed when applied to atomic systems.   
Using the Kustaanheimo-Stiefel regularization of the Coulomb potential
and introducing a fictitious time variable we have now 
made it applicable to atomic systems by using restricted GWPS and the 
time-dependent variational principle. 
The special appeal of the GWP method lies in the fact that relatively low
numbers of time-dependent basis states are sufficient to derive the spectrum 
as compared to time-independent matrix diagonalizations.
The advantage of using the fictitious time is that the computations 
are exact and analytical for the field-free hydrogen atom, which means that for 
perturbed atomic systems only the deviation of the potential from the 
Coulomb part must be taken into account in the variational approximation.
We have shown that the method can be especially adapted for systems with, 
e.g., cylindrical or spherical symmetries.

Quantum spectra of the hydrogen atom in static external fields can 
nowadays be computed quite efficiently by matrix diagonalization of
the Hamiltonian in a sufficiently large basis set, and thus the
method proposed in this paper might appear to be rather specific
as an alternative tool for studying this system with complex dynamics.
However, quantum computations for many-body Coulomb systems are certainly a 
nontrivial task.
The topic of wave packet dynamics in systems with Coulomb interactions covers
a large body of problems ranging from atomic physics to physics of solid 
state, where Coulomb interaction plays an important, often crucial, role.
In many-body physics, in particular, in the physics of solid state, 
theoretical methods well suited for studying the effects stemming from 
Coulomb interactions are still lacking.
The majority of the available methods, e.g., the method of pseudopotentials 
in atomic physics and the Fermi and the Luttinger liquid theories for solid 
conductors, are basically indirect and substantiated neither from the
theoretical nor from the experimental side.
For this reason they still remain, to a certain extent, disputable.
In the present paper we have successfully applied the Gaussian wave packet 
method to Coulomb systems with two and three nonseparable degrees of
freedom.
If the method can be further extended to larger systems with more degrees
of freedom it will allow for a wide range of future applications in 
different branches of physics.

\appendix
\section{Integrals for the diamagnetic hydrogen atom}
\label{app_integr_dia}
With the basis functions $g_m$ defined in Eq.\ \eqref{eq_KS_gwp_x_lz_m},
and using the notation $a_\mu\equiv a_\mu^k-(a_\mu^l)^*$, 
$a_\nu\equiv a_\nu^k-(a_\nu^l)^*$, $\gamma\equiv \gamma^k-(\gamma^l)^*$,
and $m\ge 0$, the absolute value of the 
magnetic quantum number the integrals in Eq.\ \eqref{eq_linGl_l_m} take the 
form
\begin{gather}
   \langle g^l_m|f(\mu,\nu)|g^k_m\rangle \nonumber \\
 = 4\pi^2\int_0^{\infty}d\mu\int_0^{\infty}d\nu (\mu\nu)^{2m+1} f(\mu^2,\nu^2)
   e^{i\left(a_\mu\mu^2+a_\nu\nu^2+\gamma\right)} \; ,
\end{gather}
where $f(\mu^2,\nu^2)$ is a poynomial in $\mu^2$ and $\nu^2$.
The integrals can be factorized, and with $x=\mu^2$ or $x=\nu^2$ the products
basically take the elementary form 
$\int_0^\infty x^n e^{-ax}dx=\frac{n!}{a^{n+1}}$ for integers $n\ge 0$ and 
${\rm Re}\, a>0$.
The integrals on the left-hand side of Eq.\ \eqref{eq_linGl_l_m} read 
\begin{eqnarray}
     \langle g^l_m|g^k_m\rangle
 &=& \frac{\pi^2(m!)^2}{(-a_\mu a_\nu)^{m+1}}
       e^{i\gamma} \equiv c \; , \nonumber \\
     \langle g^l_m|\mu^2|g^k_m\rangle
 &=& \frac{c}{-i a_\mu}(m+1) \; , \nonumber \\
     \langle g^l_m|\nu^2|g^k_m\rangle
 &=& \frac{c}{-i a_\nu}(m+1) \; , \nonumber \\
     \langle g^l_m|\mu^4|g^k_m\rangle
 &=& \frac{c}{-a_\mu^2}(m+1)(m+2) \; , \nonumber \\
     \langle g^l_m|\mu^2\nu^2|g^k_m\rangle
 &=& \frac{c}{-a_\mu a_\nu}(m+1)^2 \; , \nonumber \\
     \langle g^l_m|\nu^4|g^k_m\rangle
 &=& \frac{c}{-a_\nu^2}(m+1)(m+2) \; .
\end{eqnarray}
With the potential $V(\mu,\nu)$ given in Eq.\ \eqref{4dHsemiV}
the integrals on the right-hand side of Eq.\ \eqref{eq_linGl_l_m} are
obtained as
%
\begin{widetext}
\vskip -4ex
\begin{eqnarray}
\langle g^l_m|V(\mu,\nu)|g^k_m\rangle  & = &
 \frac{-ic}{8a_\mu^2 a_\nu^2}  
 (a_\mu+a_\nu)(1+m)[(2+3m+m^2)\beta^2+8 a_\mu a_\nu \alpha] \; , \nonumber \\
 \langle g^l_m|\mu^2 V(\mu,\nu)|g^k_m\rangle  & = & 
 \frac{c}{8 a_\mu^3 a_\nu^2}(m+1) 
 \{(1+m)(2+m)[a_\mu(2+m)+a_\nu(3+m)]\beta^2\nonumber 
 {}+8 a_\mu a_\nu[a_\mu+2a_\nu+(a_\mu + a_\nu)m]\alpha\} \; , \nonumber \\
\langle g^l_m|\nu^2 V(\mu,\nu)|g^k_m\rangle    &= &
 \frac{c}{8 a_\mu^2 a_\nu^3}(m+1) 
 \{(1+m)(2+m)[a_\mu(3+m)+a_\nu(2+m)]\beta^2\nonumber 
  {}+8 a_\mu a_\nu[2a_\mu+a_\nu+(a_\mu + a_\nu)m]\alpha\} \; .
\end{eqnarray}
%

\section{Integrals for the hydrogen atom in crossed fields}
\label{app_int_3D}
The integrals in the linear set of equations \eqref{eq_lin_Gl_3D}
take the form
\begin{gather}
   \langle g^l|f({\bf u},\nabla_{\bf u})|g^k\rangle \nonumber \\
 = \int d^4u e^{-i\left[{\bf u}(A^l)^*{\bf u}+(\gamma^l)^*\right]}
   f({\bf u},\nabla_{\bf u}) e^{i\left[{\bf u}A^k{\bf u}+\gamma^k\right]} \; .
\label{eq:B1}
\end{gather}
With the notation $A=A^k-(A^l)^*$, $\gamma=\gamma^k-(\gamma^l)^*$ 
the integrals on the left-hand side of Eq.\ \eqref{eq_lin_Gl_3D} simplify to
\begin{equation}
   I_{ij}^{lk} = \langle g^l|f_if_j|g^k\rangle
 = \int d^4u f_i f_j e^{i({\bf u}A{\bf u}+\gamma)} \; ,
\end{equation}
with $f_1=1$, $f_2=u_1^2+u_2^2$, $f_3=u_3^2+u_4^2$, $f_4=u_1u_3-u_2u_4$, and
$f_5=u_1u_4+u_2u_3$. 
The integrals have the properties $I_{ij}^{lk}=(I_{ij}^{kl})^*$ and 
$I_{ij}^{lk}=I_{ji}^{lk}$.
Using $c=\pi^2e^{i\gamma}$ and $h=1/\sqrt{-\det A}=1/(a_x^2+a_y^2-a_\mu a_\nu)$ 
we obtain
%
\begin{gather}
 I_{11}^{lk} = hc \; , \;
 I_{12}^{lk} = -i a_\nu h^2c \; , \;
 I_{13}^{lk} = -i a_\mu h^2c \; , \nonumber \\
 I_{14}^{lk} = 2i a_x h^2c \; , \;
 I_{15}^{lk} = 2i a_y h^2c \; , \;
 I_{22}^{lk} = -2 a_\nu^2 h^3c \; , \nonumber \\
 I_{23}^{lk} = -(a_\mu a_\nu+a_x^2+a_y^2) h^3c \; , \;
 I_{24}^{lk} = 4 a_\nu a_x h^3c \; , \nonumber \\
 I_{25}^{lk} = 4 a_\nu a_y h^3c \; , \;
 I_{33}^{lk} = -2 a_\mu^2 h^3c \; , \;
 I_{34}^{lk} = 4 a_\mu a_x h^3c \; , \nonumber \\
 I_{35}^{lk} = 4 a_\mu a_y h^3c \; , \;
 I_{44}^{lk} = 2 (a_y^2-3 a_x^2-a_\mu a_\nu) h^3c \; , \nonumber \\
 I_{45}^{lk} = -8 a_x a_y h^3c \; , \;
 I_{55}^{lk} = 2 (a_x^2-3 a_y^2-a_\mu a_\nu) h^3c \; .
\label{eq:B3}
\end{gather}
The integrals on the right-hand side of Eq.\ \eqref{eq_lin_Gl_3D} are
defined as 
\begin{equation}
 I_{vj}^{lk} = \langle g^l|f_j V|g^k\rangle \; .
\end{equation}\\
The potential $V$ defined via Eq.\ \eqref{regH4D} can be split into its 
harmonic and diamagnetic part,  
\begin{equation}
 V_a = \alpha{\bf u}^2 +
       \frac{1}{8}\beta^2(u_1^2+u_2^2)(u_3^2+u_4^2){\bf u}^2 \; ,
\label{eq:Va}
\end{equation}
and the terms of the paramagnetic and electric field contributions,
\begin{eqnarray}
 V_b& =& \frac{1}{2}\beta\left[(u_1p_2-u_2p_1)(u_3^2+u_4^2) \right.  \\
   && \left. +(u_3p_4-u_4p_3) (u_1^2+u_2^2)\right] + \zeta(u_1u_3-u_2u_4){\bf u}^2 \; . \nonumber 
\label{eq:Vb}
\end{eqnarray}
Note that the paramagnetic term in Eq.\ \eqref{eq:Vb} contains derivatives
with respect  
to the KS coordinates and thus the integrals must be solved by application 
of Eq.\ \eqref{eq:B1}.
We obtain
%
\begin{eqnarray}
 I_{v1a}^{lk} &=& i (a_\mu + a_\nu) [(a_\mu a_\nu + 2(a_x^2+a_y^2)) \beta^2h^2/4
            + \alpha] h^2c \; ,  \nonumber  \\
 I_{v2a}^{lk} &=& [2 a_\mu^2 a_\nu^2 + (a_x^2+a_y^2)(9a_\nu^2+2 (a_x^2+a_y^2)) + 
         a_\mu a_\nu (3 a_\nu^2 + 8 (a_x^2 + a_y^2))] \beta^2 h^5c/4  
          {} + [a_\nu (a_\mu + 2 a_\nu) + a_x^2 + a_y^2] \alpha h^3c \;, \nonumber  \\
 I_{v3a}^{lk} & = & [3 a_\mu^3 a_\nu + 8 a_\mu a_\nu (a_x^2 + a_y^2) + 
                 2 (a_x^2 + a_y^2)^2 + a_\mu^2 (2 a_\nu^2 + 9 (a_x^2
                 + a_y^2))] \beta^2 h^5c/4 
{} + [a_\mu (2 a_\mu + a_\nu) + a_x^2 + a_y^2] \alpha h^3c \; , \nonumber  \\
 I_{v4a}^{lk} &=& -(a_\mu + a_\nu) a_x [3 (a_\mu a_\nu + a_x^2 + a_y^2)\beta^2h^2
                 + 4\alpha] h^3c \; , \nonumber  \\
 I_{v5a}^{lk} &=& -(a_\mu + a_\nu) a_y [3 (a_\mu a_\nu + a_x^2 + a_y^2)\beta^2h^2
                 + 4\alpha] h^3c \; ,  \label{eq_rechte_seite1} 
\end{eqnarray}
%
\begin{eqnarray}
 I_{v1b}^{lk} & = & 2(a_\mu+a_\nu)(a_y^k a_x \beta-a_x^k a_y \beta+a_x \zeta)
               h^3c \; , \nonumber \\
 I_{v2b}^{lk} & = & -2i (2 a_\mu a_\nu + 3 a_\nu^2 + a_x^2 + a_y^2) 
              (a_y^k a_x \beta - a_x^k a_y \beta + a_x \zeta) h^4c \; , \nonumber \\
 I_{v3b}^{lk} & = & -2i (3 a_\mu^2 + 2 a_\mu a_\nu + a_x^2 + a_y^2) 
              (a_y^k a_x \beta - a_x^k a_y \beta + a_x \zeta) h^4c \; , \nonumber \\
 I_{v4b}^{lk} & = & 2i (a_\mu + a_\nu) [-6 a_x^k a_x a_y \beta + 
               a_y^k (a_\mu a_\nu + 5 a_x^2 - a_y^2) \beta + 
               (a_\mu a_\nu + 5 a_x^2 - a_y^2) \zeta] h^4c \; , \nonumber \\
 I_{v5b}^{lk} & = & -2i (a_\mu + a_\nu) 
              [a_x^k (a_\mu a_\nu - a_x^2 + 5 a_y^2) \beta - 
              6 a_x a_y (a_y^k \beta + \zeta)] h^4c \; ,
\label{eq_rechte_seite2}
\end{eqnarray}
\end{widetext}
where $a_x^k$ and $a_y^k$ are elements of the width matrix $A^k$.
The right-hand side vector in Eq.\ \eqref{eq_lin_Gl_3D} is the sum of two 
corresponding terms in Eqs.\ \eqref{eq_rechte_seite1} and 
\eqref{eq_rechte_seite2}, i.e.,
$I_{vj}^{lk}=I_{vja}^{lk}+I_{vjb}^{lk}$ for $j=1,\dots,5$.

\section{Structure of the matrices $B$ and $C$}
\label{app_param_red}
A structure of the matrices $B^k$ and $C^k$ is searched which is preserved
in the matrix product $V_2^kC^k$ in Eq.\ \eqref{eq_BC_motion},
where $V_2^k$ has the same structure as $A$ in Eq.\ \eqref{eq:A_sym}.
This is provided by the form
%
\begin{equation}
 B  =  \left(
 \begin{array}{rrrr}
 b_{11} &  b_{12} &  b_{13} &  b_{14} \\
-b_{12} &  b_{11} &  b_{14} & -b_{13} \\
 b_{31} &  b_{32} &  b_{33} &  b_{34} \\
 b_{32} & -b_{31} & -b_{34} &  b_{33} 
 \end{array} \right) , \nonumber 
\end{equation}
\begin{equation}
C = \left(
 \begin{array}{rrrr}
  c_{11} &  c_{12} &  c_{13} &  c_{14} \\
 -c_{12} &  c_{11} &  c_{14} & -c_{13} \\
  c_{31} &  c_{32} &  c_{33} &  c_{34} \\
  c_{32} & -c_{31} & -c_{34} &  c_{33} 
\end{array} \right) ,
\label{eq_BC_sym}
\end{equation}
as can easily be shown by explicit multiplication.
The superscript $k$ running over all GWPs has been omitted here.
Compared to Eq.\ \eqref{eq:A_sym} the number of independent parameters per 
matrix increases from 4 to 8, however, this is still less than 16 parameters 
for a general $4\times 4$ matrix without any special structure.

The matrix $A=\frac{1}{2}BC^{-1}$ can be calculated analytically.
To this end we introduce the auxiliary matrix
\newpage
\begin{equation}
 D = \left(
\begin{array}{rrrr}
  c_{33} & -c_{34} & -c_{13} & -c_{14} \\
  c_{34} &  c_{33} & -c_{14} &  c_{13} \\
 -c_{31} & -c_{32} &  c_{11} & -c_{12} \\
 -c_{32} &  c_{31} &  c_{12} &  c_{11} 
\end{array} \right) .
\label{eq_D_sym}
\end{equation}
%
The product $C_1=CD$ yields 
\begin{gather}
 C_1 = \left(
\begin{array}{rrrr}
  h & k & 0 &  0 \\
 -k & h & 0 &  0 \\
  0 & 0 & h & -k \\
  0 & 0 & k &  h 
\end{array} \right) , \\ \text{with} \; \left\{
\begin{array}{rr}
 h =& -c_{13}c_{31} - c_{14}c_{32} + c_{11}c_{33} + c_{12}c_{34} \; , \\
 k =&  c_{14}c_{31} - c_{13}c_{32} + c_{12}c_{33} - c_{11}c_{34} \; .
\end{array} \right. \nonumber 
\end{gather}
The matrix $C_1$ can be easily inverted, and thus allows for the
calculation of $A=\frac{1}{2}BC^{-1}=\frac{1}{2}BDC_1^{-1}$.
With $h'\equiv h/[2(h^2+k^2)]$, $k'\equiv k/[2(h^2+k^2)]$ the four 
independent parameters in Eq.\ \eqref{eq:A_sym} read
\begin{eqnarray}
 a_\mu &=& (b_{11}c_{33} - b_{14}c_{32} + b_{12}c_{34} - b_{13}c_{31})h' \nonumber \\
       &&+ (b_{14}c_{31} + b_{12}c_{33} - b_{11}c_{34} - b_{13}c_{32})k' \; ,
 \nonumber \\
 a_\nu &=& (b_{33}c_{11} + b_{34}c_{12} - b_{31}c_{13} - b_{32}c_{14})h' \nonumber \\
      &&   + (b_{33}c_{12} - b_{34}c_{11} - b_{32}c_{13} + b_{31}c_{14})k' \; ,
 \nonumber \\
 a_x  &=& (b_{13}c_{11} + b_{14}c_{12} - b_{11}c_{13} - b_{12}c_{14})h' \nonumber \\
      &&  + (b_{13}c_{12} - b_{14}c_{11} - b_{12}c_{13} + b_{11}c_{14})k' \; ,
 \nonumber \\
 a_y  &=& (b_{14}c_{11} - b_{13}c_{12} + b_{12}c_{13} - b_{11}c_{14})h' \nonumber \\
      &&  + (b_{13}c_{11} + b_{14}c_{12} - b_{11}c_{13} - b_{12}c_{14})k' \; .
\end{eqnarray}
%


\end{document}